\documentclass[twocolumn]{autart}    

\newtheorem{assumption}{Assumption}
\newtheorem{theorem}{Theorem}

\newtheorem{lemma}{Lemma}
\newtheorem{definition}{Definition}
\newtheorem{example}{Example}
\newtheorem{corollary}{Corollary}
\usepackage{times} 
\usepackage{graphics} 
\usepackage{epsfig} 
\usepackage{mathptmx} 
\usepackage{amsmath} 
\usepackage{amssymb}  
\usepackage{subfigure}

\usepackage{color}
\usepackage{xcolor}

\newcommand{\sys}[4]{\mbox{$\left[
\begin{array}{c|c}
#1 & #2 \\ \hline #3 & #4
\end{array}
\right]$}}
\usepackage{graphicx}          

\begin{document}

\begin{frontmatter}

\title{Mean-square Stabilizability via Output Feedback for Non-minimum Phase Networked Feedback Systems \thanksref{footnoteinfo}}

\thanks[footnoteinfo]{Corresponding author W. Su, Tel. +86-20-87111804.
Fax +86-20-87114612.}

\author[WSu]{Weizhou Su}\ead{wzhsu@scut.edu.cn},    
\author[WSu]{Jieying Lu}\ead{aujylu@scut.edu.cn},
\author[WSu]{Yilin Wu}\ead{lyw@gdei.edu.cn},               
\author[MFu]{Minyue Fu}\ead{minyue.fu@newcastle.edu.au}, 
\author[JChen]{Jie Chen}\ead{jichen@cityu.edu.hk}  

\address[WSu]{School of Automation Science and Engineering, South China University of Technology}
\address[MFu]{School of Electrical and Computer Engineering, University of Newcastle} 
\address[JChen]{Department of Electronic Engineering, City University of Hong Kong}   


\begin{abstract}                          
This work studies mean-square stabilizability via output feedback for a networked linear time invariant (LTI) feedback system with a non-minimum phase plant. In the feedback system, the control signals are transmitted to the plant over a set of parallel communication channels with possible packet dropout. Our goal is to analytically describe intrinsic constraints among channel packet dropout probabilities and the plant's characteristics, such as unstable poles, non-minimum phase zeros in the mean-square stabilizability of the system. It turns out that this is a very hard problem. Here, we focus on the case in which the plant has relative degree one and each non-minimum zero of the plant is only associated with one of control input channels. Then, the admissible region of packet dropout probabilities in the mean-square stabilizability of the system is obtained. Moreover, a set of hyper-rectangles in this region is presented in terms of the plant's non-minimum phase zeros, unstable poles and Wonham decomposition forms which is related to the structure of controllable subspace of the plant. When the non-minimum phase zeros are void, it is found that the supremum of packet dropout probabilities' product in the admissible region is determined by the product of plant's unstable poles only. A numerical example is presented to illustrate the fundamental constraints in the mean-square stabilizability of the networked system.
\end{abstract}

\end{frontmatter}

\section{Introduction}
In the last two decades, stabilization problems for networked feedback systems have attracted a great amount of research interests (for example, see \cite{05FX}, \cite{03IF}, \cite{02NE}, \cite{07NFSE}, \cite{14VCS} and the references therein). These works mainly focus on coping with new challenges caused by limited resources, uncertainties/unreliability in communication channels. Indeed, great success has been achieved in this research area, in particular, for stabilization via state feedback. In \cite{05E}, networked multi-input multi-output (MIMO) LTI feedback systems are studied where control signals are transmitted to actuators over fading channels. Uncertainties in the channels are modeled as multiplicative noises and then a design scheme is presented for mean-square stabilization via state feedback. Moreover, fundamental constraints in mean-square stabilizability caused by channel uncertainties are studied for the networked systems in \cite{05E}. It is shown for a networked single-input feedback system that the minimum capacity required for mean-square stabilization via state feedback is determined by the product of all the unstable poles of the plant. In \cite{09XXQ}, this problem is studied for a networked MIMO system where the total capacity of the feedback control channels is given and can be allocated to each channels associated with individual control inputs. It is found that the minimum total channel capacity for the mean-square stabilization problem is also determined by the product of all the unstable poles of the plant.
Some new developments in stabilization and state estimation for networked systems over packet dropping channels are presented in \cite{11EE} for systems with both actuators and sensors connected to controllers over communication channels.

In this work, we study the mean-square stabilizability via output feedback for a networked MIMO LTI system
where the control signals are transmitted over packet dropping channels. The channel uncertainties are also
modeled as multiplicative noises. The difficulties for mean-square stabilization with multiplicative noises
are well recognized (see e.g. \cite{02LS}), especially for the case with non-minimum phase zeros
\cite{14QCSF}. Here, we attempt to explore fundamental constraints among channel packet dropout probabilities and plant's characteristics and structure in mean-square stabilizability of the networked system with a non-minimum phase plant. With this purpose, our study focuses on the case in which the plant is with relative degree one and each non-minimum phase zero is associated with one of control input channels.
The largest admissible region of packet dropout probabilities for mean-square stabilizability of the system is presented. Moreover, a set of hyper-rectangles in this region is found in terms of plant's nonminimum phase zeros, unstable poles and Wonham decomposition forms \cite{67W}. The boundaries of these hyper-rectangles describe the interactions between channel packet dropout probabilities and the plant's characteristics and structure in this problem. Moreover, to explain the features of this admissible region comprehensively, we introduce a concept, blocking packet dropout probability with which data transmitted over all channels are lost. An upper bound of this probability allowed to the mean-square stabilizability is presented for the non-minimum phase networked system. In particular, it is shown that when non-minimum phase zeros of the plant are void, the supremum of blocking packet dropout probability in the admissible region is determined by the product of all the unstable poles of the plant. We note that for the latter case, a related problem was studied under certain technical assumption in \cite{09XXQ}, but this assumption is removed in this work.

The remainder of this paper is organized as follows. We proceed in Section 2 to formulate the problem under
study. A useful tool, upper triangular coprime factorization, is developed in Section 3. Section 4 presents our main results on mean-square stabilizability via output feedback for the networked systems. Section 5 concludes the paper.

The notation used throughout this paper is fairly standard. For any complex number $z$, we denote its
complex conjugate by $\bar{z}$. For any vector $u$, we denote its transpose by $u^T$ and conjugate transpose by $u^*$. For any matrix $A$, the transpose, conjugate transpose, spectral radius and trace are denoted by $A^T$, $A^*$, $\rho(A)$ and $\mbox{Tr}(A)$, respectively. Denote a state-space model of an LTI system by $\sys{A}{B}{C}{D}$. For any real rational function matrix $G(z)$, $z\in \mathbb{C}$, define
$G^{\sim}(z)=G^T(1/z)$. Denote the expectation operator by $\mathbf{E}\{\cdot\}$. Let the open unit disc be
denoted by $\mathbb{D}:=\{z\in\mathbb{C}:|z|<1\}$, the closed unit disc by
$\bar{\mathbb{D}}:=\{z\in\mathbb{C}:|z|\leq1\}$, the unit circle by
$\partial\mathbb{D}$, and the complements of $\mathbb{D}$ and
$\bar{\mathbb{D}}$ by $\mathbb{D}^c$ and $\bar{\mathbb{D}}^c$, respectively. The space $\mathcal{L}_2$ is a Hilbert space and consists of all complex matrix functions $G(z)$ which are measurable in $\partial\mathbb{D}$ and
\begin{align*}
\frac{1}{2\pi}\int^{\pi}_{-\pi}{\rm Tr}\left[G^*(e^{j\theta})G(e^{j\theta})\right]
d\theta <\infty.
\end{align*}
For $F, G \in \mathcal{L}_2$, the inner product is defined as
\begin{align}\label{inner_product}
\langle F,G\rangle
=\frac{1}{2\pi}\int^{\pi}_{-\pi}\mbox{Tr}\left[F^*(e^{j\theta})G(e^{j\theta})\right]d\theta
\end{align}
and the induced norm is defined by
\begin{align}\label{induced_norm}
\|G\|_2=\sqrt{\langle G,G\rangle}.
\end{align}
It is well-known that $\mathcal{L}_2$ admits an orthogonal
decomposition into the subspaces
\begin{align*}
\mathcal{H}_2:=&\Bigg\{G : G(z)~\text{analytic
in}~\bar{\mathbb{D}}^c, \\
&\hspace{-0.8cm}\|G\|_2
=\left\{\sup\limits_{r>1}\frac{1}{2\pi}\int^{\pi}_{-\pi}{\rm Tr}\left[G^*(re^{j\theta})G(re^{j\theta})\right]
d\theta\right\}^{\frac{1}{2}}<\infty\Bigg\},
\end{align*}
and
\begin{align*}
\mathcal{H}_2^{\perp}:=&\Bigg\{G : G(z)~\text{analytic
in}~\mathbb{D},~G(0)=0,\\
&\hspace{-0.8cm}\|G\|_2
=\left\{\sup\limits_{r<1}\frac{1}{2\pi}\int^{\pi}_{-\pi}{\rm Tr}\left[G^*(re^{j\theta})G(re^{j\theta})\right]
d\theta\right\}^{\frac{1}{2}}<\infty\Bigg\}.
\end{align*}
Note that for any $F\in\mathcal{H}^{\perp}_2$ and  $G\in\mathcal{H}_2$,
\begin{align}\label{orth}
\langle F, G\rangle=0.
\end{align}
Define the Hardy space
\[
\mathcal{H}_{\infty}:=\left\{G : G(z)~\text{bounded and analytic
in}~\mathbb{D}^c\right\}.
\]
A subset of $\mathcal{H}_{\infty}$, denoted by
$\mathcal{RH}_{\infty}$, is the set of all proper stable rational transfer function matrices in the discrete-time
sense. Note that we have used the same notation $\|\cdot\|_2$ to denote the corresponding norm for spaces
$\mathcal{L}_2$, $\mathcal{H}_2$ and $\mathcal{H}_2^{\perp}$.

\section{Problem Formulation}

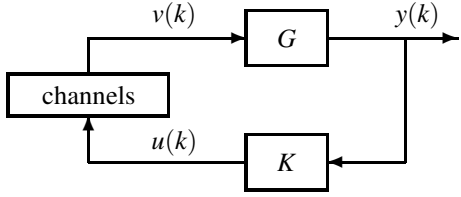
\begin{figure}[hbt]
\thicklines \centering
\begin{picture}(300,65)
\put(60,52){\vector(1,0){60}}\put(120,42){\framebox(30,20){$G$}}
\put(150,52){\vector(1,0){50}}
\put(180,52){\line(0,-1){47}}
\put(180,5){\vector(-1,0){30}}
\put(120,5){\line(-1,0){60}}
\put(30,23){\framebox(60,15){channels}}\put(120,-5){\framebox(30,20){$K$}}
\put(60,38){\line(0,1){14}}
\put(60,5){\vector(0,1){18}}
\put(90,10){\makebox(5,5){$u(k)$}}
\put(182,58){\makebox(5,5){$y(k)$}}\put(90,58){\makebox(5,5){$v(k)$}}
\end{picture}
\caption{A networked feedback system}
\label{setupfeedback}
\end{figure}

The networked feedback system under study is depicted in Fig.
\ref{setupfeedback}. The plant $G$ in the system is a MIMO LTI system and the signal $y(k)$ is the
measurement. The control signal $u(k)$ for the plant is generated by the feedback controller $K$. It includes $r$ entries $u_1(k), \cdots, u_r(k)$ which are sent
to the plant $G$ over $r$ parallel packet dropping channels, respectively. The signal $v(k)=[ v_1(k), \cdots, v_r(k)]^T$ is the received control signal at the plant side.

Let $\left\{\alpha_j(k), k=0,1,2,\cdots,\infty \right\}$, $j=1,\cdots,r$ be random
processes with independent identical Bernoulli probability distributions, respectively. It indicates the receipt of the control signal $u(k)$, i.e., $\alpha_j(k)=1$ if $u_j(k)$ is received, otherwise $\alpha_j(k)=0$. Let the probability of $\alpha_j(k)=0$ be $p_j$. The averaged receiving rate of data packets is $\mathbf{E}\{\alpha_j(k)\}=1-p_j$ in the $j$-th channel.  Let $\omega_j(k)=\alpha_j(k)-(1-p_j)$. Subsequently, the received control signal $v_j(k)$ is written as:
\begin{align}\label{j_fading_channel}
v_j(k)=\alpha_j(k)u_j(k)=(1-p_j)  u_j(k)+\omega_j(k)u_j(k).
\end{align}
It is clear that $\left\{\omega_j(k), k=0,1,2,\cdots,\infty \right\}$, $j=1,\cdots,r$ have independent identical probability distributions, referred to as {\em i.i.d} random processes, respectively. The {\em i.i.d} random
process $\left\{\omega_j(k), k=0,1,2,\cdots,\infty \right\}$ has zero mean and variance $(1-p_j)p_j$.
Now, it is assumed that $\{\alpha_j(k)\}$, $j=1,\cdots,r$ are mutually independent. And then, it holds for any $i, j \in \left\{1, \cdots, r\right\}$, $i\neq j$ that $\mathbf{E}\{\omega_i(k_1)\omega_j(k_2)\}=0, \forall k_1,k_2>0.$

Denote the averaged channel gain by
\begin{align*}
\mu={\rm diag}\left\{1-p_1, \cdots, 1-p_r \right\}
\end{align*}
and the multiplicative noise in the channels by
\begin{align}\label{omega}
\omega(k)={\rm diag}\left\{\omega_1(k), \cdots, \omega_r(k) \right\}.
\end{align}
It follows from the discussion above that  ${\bf E}(\omega(k))=0$  and
$$
{\bf E}(\omega(k)\omega^T(k))={\rm diag}\{p_1(1-p_1), \cdots, p_r(1-p_r) \}.
$$
Let $\bar{\omega}(k)=\mu^{-1}\omega(k)$. From (\ref{j_fading_channel}), the packet dropout channels in the system shown in Fig. \ref{setupfeedback} are modeled as follows (also see \cite{05E}):
\begin{align}\label{fading_channel}
v(k)=\mu u(k)+\mu \bar{\omega}(k) u(k).
\end{align}
It is verified from the mean and covariance of $\omega(k)$, $k=0,1,2,\cdots$ that
\begin{align*}
{\bf E}(\bar{\omega}(k))=0
\quad\;\;  {\rm and} \quad\;\;
{\bf E}(\bar{\omega}(k)\bar{\omega}^T(k))=\Sigma
\end{align*}
where
$$\Sigma={\rm diag}\left\{{\displaystyle \frac{p_1}{1-p_1}},\cdots,{\displaystyle \frac{p_r}{1-p_r}}\right\}.$$

\begin{definition} (see \cite{71WB})
For any initial state, if it holds for the control signal
and the output that
\[
\underset{k \rightarrow \infty}\lim \mathbf{E}\left\{u(k)u^T(k)\right\}=0, \quad\quad
\underset{k \rightarrow \infty}\lim \mathbf{E}\left\{y(k)y^T(k)\right\}=0,
\]
then the feedback system in Fig. \ref{setupfeedback} is said to be mean-square stable.
\end{definition}

\begin{figure}[hbt]
\thicklines \centering
\begin{picture}(180,60)
\put(70,40){\framebox(30,20){$T$}}
\put(40,50){\vector(1,0){30}}
\put(100,50){\line(1,0){30}}
\put(130,10){\line(0,1){40}}
\put(130,10){\vector(-1,0){30}}
\put(70,0){\framebox(30,20){$\bar{\omega}$}}
\put(70,10){\line(-1,0){30}}
\put(40,10){\line(0,1){40}}
\put(20,30){\makebox(5,5){$\Delta(k)$}}
\put(145,30){\makebox(5,5){$u(k)$}}
\end{picture}
\caption{An LTI system with a multiplicative noise}
\label{setupfeedback2}
\end{figure}
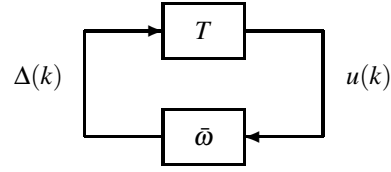

To study the mean-square stability for the networked feedback system in Fig. \ref{setupfeedback}, it is re-diagrammed as an LTI  system with a multiplicative noise as shown in Fig. \ref{setupfeedback2}. Let
$\Delta(k)=\bar{\omega}(k) u(k).$ The channel model (\ref{fading_channel}) is rewritten as
\begin{align*}
v(k)=\mu u(k)+\mu \Delta(k).
\end{align*}
Thus, the transfer function $T$ from $\Delta(k)$ to $u(k)$ in the nominal system is given by
\begin{align}\label{T}
T=(I-KG\mu)^{-1}KG\mu
\end{align}
where $G\mu$ is considered as a new plant involved with the averaged gain of the channel.
Let $T_{ij}$, $i,j=1,\cdots,r$ be the $\{i,j\}$-th entry of the transfer function matrix $T$ and
\begin{align}\label{T_hat}
\hat{T}=\begin{bmatrix}
\|T_{11}\|_2^2  &  \cdots & \|T_{1r}\|_2^2 \\
& \cdots &   \\
\|T_{r1}\|_2^2  &  \cdots & \|T_{rr}\|_2^2
\end{bmatrix}
.
\end{align}
\begin{lemma}(see \cite{02LS})\label{Mean_square_stable}
The LTI system with a multiplicative noise in Fig. \ref{setupfeedback2} is mean-square stable if and only it
holds that
\begin{align}\label{speactral_radius_MS_stable}
\rho(\hat{T}\Sigma)<1.
\end{align}
\end{lemma}

To design an output feedback controller $K$ which stabilizes the system in Fig. \ref{setupfeedback2} in the mean-square sense is referred to as mean-square stabilization via output feedback. If this problem is solvable, the system is refereed to as mean-square stabilizable. Intuitively, the mean-square stabilizability of the system is related to the packet dropout probabilities $p_1$, $\cdots$, $p_r$ and the transfer function $T$
of the nominal closed-loop system. In this work, fundamental constraints in mean-square stabilizability
via output feedback are studied for the networked system in terms of the packet dropout probabilities and characteristics of the plant $G$.

\section{Upper Triangular Coprime Factorization}\label{Upper_Tri_CPF}

To study the mean-square stabilizability of the networked system, we consider the set of all possible stabilizing controllers for  the plant $G\mu$, which is described by Youla parametrization in terms of its coprime factorizations. A useful tool for the mean-square stabilization design, referred to as upper triangular coprime factorization, is introduced in this section.

Suppose that the state-space model of the plant $G\mu$ is given by $G\mu=\sys{A}{B}{C}{0}$, and $\{A,B\}$ is controllable, $\{A, C\}$ is detectable. Let the right coprime factorization of the plant $G\mu$ be $NM^{-1}$,
where the factors $N$ and $M$ are from $\mathcal{RH}_{\infty}$. Moreover, $N$ and $M$ are given by
\begin{align}
&M=I-F(zI-A+BF)^{-1}B\label{State_Space_M},\\
&N=C(zI-A+BF)^{-1}B\label{State_Space_N},
\end{align}
where $F$ is any stabilizing state feedback gain (for details, see e.g. \cite{96ZDG}).

It is shown in \cite{67W} that, with certain state transformation, the state-space model of $G\mu$ can be
transformed into so-called Wonham decomposition form $\sys{A_w}{B_w}{C_w}{0}$ with
\[
A_w=\begin{bmatrix} A_1    & \star     & \cdots & \star     \\ 0      & A_2   & \cdots & \star     \\
\vdots &\vdots & \ddots & \vdots \\
0      & 0     & \cdots & A_r
\end{bmatrix},
\quad  B_w=\begin{bmatrix}
b_1    & \star     & \cdots & \star    \\ 0      & b_2   & \cdots & \star     \\
\vdots &\vdots & \ddots & \vdots \\
0      & 0     & \cdots & b_r
\end{bmatrix},
\]
where
\begin{align}\label{jj_pair}
A_j=\begin{bmatrix}
0  &  1  &  0 & \cdots &  0 \\
0  &  0  &  1 & \cdots &  0 \\
&\cdots  &  \cdots &  \\
0  &  0  &  0 & \cdots &  1 \\
-a_{j l_j} &-a_{j(l_j-1)} &-a_{j(l_j-2)} & \cdots & -a_{j 1}
\end{bmatrix},
\quad b_j=\begin{bmatrix}
0\\
0\\
\vdots\\
0\\
1
\end{bmatrix}
.
\end{align}
Since the pairs $\{A_j, b_j\}$, $j=1,\cdots,r,$ are all controllable, it is always
possible to find row vectors $f_j$ such that $A_j+b_jf_j$ is stable for all $j=1,\cdots,
r$. Now, we select a block diagonal state feedback gain
$F={\rm diag}\left\{ f_1, f_2, \cdots, f_r \right\}$.
Applying Wonham decomposition forms and the state feedback gain $F$ into (\ref{State_Space_M}) and
(\ref{State_Space_N}) yields a right coprime factorization $G\mu=NM^{-1}$ in which the factor $M$ is an
upper triangular matrix. In this work, this coprime factorization is referred to as {\em upper triangular
coprime factorization}. It is summarized in the following result.
\begin{lemma}\label{UT}
For a given plant $G\mu$, there exist coprime matrices $N$ and $M\in \mathcal{RH}_{\infty}$ such that
$G\mu=NM^{-1}$ and the matrix $M$ is an upper triangular matrix. Furthermore, the diagonal elements $m_{jj}$, $j=1,\cdots, r$ of $M$ are given by
\begin{align*}
m_{jj}&=1-f_j(zI-A_j+b_jf_j)^{-1}b_j,\quad j=1,\cdots,r.
\end{align*}
\end{lemma}
Taking account of the structures of $A_j$ and $b_j$, we can see that the numerator polynomial of $m_{jj}$ is the
characteristic polynomial of $A_j$. Denote the unstable poles of $A_j$ by $\lambda_{j 1}, \cdots, \lambda_{j l_j}$. Note the fact that $\left\{A_j,b_j\right\}$ is controllable. By selecting a proper $f_j$, the poles of $m_{jj}$ are assigned as $1/\lambda_{j 1},\cdots, 1/\lambda_{j l_j}$ and all stable poles of $A_j$.
This yields that the diagonal elements $m_{jj}$ is given by
$m_{jj}={\displaystyle \frac{(z-\lambda_{j 1})\times \cdots \times (z-\lambda_{j l_j})}{(\lambda_{j 1}^*z-1)\times \cdots
\times (\lambda_{j l_j}^*z-1)}}.$
It is clear that $m_{jj}$ is an inner, i.e., $m^{\sim}_{jj}(z)m_{jj}(z)=1$ (for
definition of an inner, see e.g. \cite{96ZDG}). Denote it by $m_{j,in}$. For this particular upper triangular coprime factorization, let $M_{in}={\rm diag}\left\{m_{1, in} \cdots, m_{r, in}\right\}$ referred to as diagonal inner.
Moreover, a balanced realization of $m_{j, in}$, which is used in remainder of this work, is denoted by
\begin{align}
m_{j, in}=\sys{A_{j,in}}{B_{j, in}}{C_{j, in}}{D_{j, in}}.
\end{align}

{In general, for a given plant $G$, there is a finite number of Wonham decomposition forms to $G\mu$ in which poles of the plant could be assigned to different diagonal sub-matrixes in the state matrix $A_w$, respectively. This comes out a set of upper triangular coprime factorizations and associated diagonal inners $M_{in}$ for the plant, which are dependent to unstable poles in diagonal sub-matrixes in Wonham decomposition forms. It will be shown in next section that the interaction between this feature and non-minimum phase zeros of the plant leads to the non-convexity in analyzing the mean-square stabilizability for the non-minimum phase system.

\section{Mean-square stabilizability}

In this section, fundamental constraints in mean-square stabilizability via output feedback, caused by the uncertainties in network channels, are studied for the system in Fig. \ref{setupfeedback}. This is a very hard problem in general since non-minimum phase zeros make the mean-square stabilization via output feedback
to be a non-convex problem (see for example \cite{14QCSF}).  Our study focuses on a non-minimum phase plant under Assumption 1.
\begin{assumption}\label{NMP_zero}
The plant $G$ has non-minimum phase zeros $z_1, \cdots, z_r$. Each of them is associated with a column of $G$,
i.e.
$$
G=G_0{\rm diag}\left\{ 1-z_1z^{-1}, \cdots, 1-z_rz^{-1}\right\}
$$
where $G_0$ is a minimum phase system  and with relative degree one, i.e.,
${\displaystyle \underset{|z|\rightarrow \infty}\lim zG_0(z)}$ is invertible.
\end{assumption}
At first glance, this assumption is quite artificial. However, due to multi-path transmission in wireless communication, multiple paths with different propagation lengths yield a channel with finite impulse response (FIR) which may include a nonminimum phase zero.
In general, there is as called ``common sub-channel zero" induced by multi-path transmission which is a difficult issue in channel identification and estimation (for example see \cite{03LD} and \cite{95T}). This is a case which fits Assumption 1. On the other hand, we attempt to analytically investigate inherent constraints on the mean-square stabilizability imposed by interaction between Wonham decomposition forms and non-minimum pase zeros of the plant for the networked system. To seek a simplicity, the plants under this assumption are studied, which would be an interesting case as shown in Example 1. It should be noted that the results in this work can be extended to the case:
$$
G=G_0{\rm diag}\left\{ z^{-\tau_1}g_1, \cdots, z^{-\tau_r}g_r\right\}
$$
where scale transfer functions $g_j$, $j=1,\cdots, r$ have more than one non-minimum phase zeros and relative degree zero, $\tau_j$, $j=1,\cdots, r$ are positive integers, $G_0$ is a minimum phase system  and with relative degree one.

Now, we consider all stabilizing controllers for the nominal closed-loop system $T$. Let $NM^{-1}$ be an upper triangular right coprime factorization of the plant $G\mu$, which is discussed
in the preceding section. And let $\tilde{M}^{-1}\tilde{N}$, with $\tilde{M}$,
$\tilde{N}\in \mathcal{RH}_{\infty}$, be the left coprime factorization of
the plant $G\mu$ associated with $NM^{-1}$. It is well known (see \cite{96ZDG} for details)
that the factors $N$, $M$, $\tilde{N}$, $\tilde{M}$ with some $X$, $Y$, $\tilde{X}$,
$\tilde{Y}\in \mathcal{RH}_{\infty}$ satisfy the Bezout Identity below:
\begin{align}\label{bezout}
\begin{bmatrix}
M&Y\\ N&X
\end{bmatrix}
\begin{bmatrix}
\tilde{X}&-\tilde{Y}\\
-\tilde{N}&\tilde{M}
\end{bmatrix}
=I.
\end{align}
All stabilizing controllers for the nominal system are given
\begin{align}\label{YouLa}
K=(\tilde{X}-Q\tilde{N})^{-1}(\tilde{Y}-Q\tilde{M}),
\end{align}
where $Q\in \mathcal{RH}_{\infty}$ is a parameter to be designed. Applying the controller (\ref{YouLa}) to the system, we  obtain the nominal closed-loop system $T$ in (\ref{T}) as follows:
\begin{align}\label{TF_R}
T=(Y-MQ)\tilde{N}.
\end{align}
According to Lemma \ref{Mean_square_stable}, the system is
mean-square stabilizable if and only if there exists a $Q$ satisfying the inequality $\rho(\hat{T}\Sigma)<1$.

To this end, we need the following result (see \cite{85HJ} for details),
\begin{lemma}\label{Spectral_radius_Lemma}
Suppose  $W$ is an $r \times r$ positive matrix and $w_{ij}$ is the $\{i,j\}$-th entry of $W$. Then, it
holds that
\begin{align*}
\rho(W)=\underset{\Gamma}{\inf}\;\underset{j}{\max}\sum_{i=1}^r\frac{\gamma_i^2}{\gamma_j^2} w_{ij}
\end{align*}
where $\Gamma={\rm diag}\left\{\gamma_1^2, \cdots, \gamma_r^2\right\}$, with $\gamma_i> 0$, $i=1,\cdots,r$.
\end{lemma}

Denote the $j$-th column of $T$ by $T_j$. Applying Lemma \ref{Spectral_radius_Lemma},
we have
\begin{align}\label{Spectral_radius_T1}
\rho(\hat{T}\Sigma)=\underset{\Gamma}{\inf}\;\underset{j}{\max} \left\|\Gamma^{1/2} T_j
\gamma_j^{-1}\right\|_2^2\frac{p_j}{1-p_j}.
\end{align}
From Lemma \ref{Mean_square_stable} and the spectral radius given in (\ref{Spectral_radius_T1}),
we have the next result.
\begin{lemma}\label{lemma_stable}
The closed-loop system in Fig.~ \ref{setupfeedback2} is mean-square stabilizible if and only if it holds for some $\Gamma$ and $Q$ that
\begin{align}\label{H_c}
\left\|\Gamma^{1/2} T_j \gamma_j^{-1}\right\|_2^2\frac{p_j}{1-p_j}<1, \quad j=1,\cdots,r.
\end{align}
\end{lemma}

Now, it is studied to minimize $\rho(\hat{T}\Sigma)$.
From (\ref{TF_R}), it holds for the system that
\begin{align}\label{TF_Rj}
\Gamma^{1/2} T_j \gamma_j^{-1}=\Gamma^{1/2}(Y-MQ)\tilde{N}\Gamma^{-1/2}e_j
\end{align}
where $e_j$ is the $j$-th column of the $r \times r$ identity matrix $I$.
Applying Bezout identity (\ref{bezout}) into (\ref{TF_Rj}) leads to
\begin{align}\label{TF_RjA}
\Gamma^{1/2} T_j \gamma_j^{-1}=\Gamma^{1/2}[M(\tilde{X}-
Q \tilde{N})-I]\Gamma^{-1/2}e_j.
\end{align}
Let $M_{\Gamma}=\Gamma^{1/2} M \Gamma^{-1/2}$, $\tilde{N}_{\Gamma}=\Gamma^{1/2} \tilde{N} \Gamma^{-1/2}$, $\tilde{X}_{\Gamma}=\Gamma^{1/2} \tilde{X} \Gamma^{-1/2}$, and $Q_{\Gamma}=\Gamma^{1/2} Q \Gamma^{-1/2}$. We rewrite (\ref{TF_RjA}) as
\begin{align*}
\Gamma^{1/2} T_j \gamma_j^{-1}=[M_{\Gamma}(\tilde{X}_{\Gamma}-
Q_{\Gamma} \tilde{N}_{\Gamma})-I]e_j.
\end{align*}
Let the inner-outer factorization of  $M_{\Gamma}$ given by $M_{\Gamma}=M_{\Gamma in}M_{\Gamma out}$ where $M_{\Gamma in}$, $M_{\Gamma out}$ are inner and outer,
respectively (see e.g. \cite{96ZDG}). Noting the identity $M_{\Gamma in}^{\sim}M_{\Gamma in}=I$ and the definition of ${\mathcal L}_2$ norm, we have that
\begin{align}\label{Inner_factorization1}
\hspace{0cm}\left\|\Gamma^{1/2}T_j \gamma_j^{-1}\right\|_2^2
=\left\|\left[M_{\Gamma out}(\tilde{X}_{\Gamma}- Q_{\Gamma} \tilde{N}_{\Gamma})-M_{\Gamma
in}^{-1}\right]e_j\right\|_2^2.
\end{align}
Due to the facts that $M_{\Gamma in}^{-1}-M_{\Gamma in}^{-1}(\infty) \in \mathcal{H}_2^{\perp}$ and $M_{\Gamma
out}(\tilde{X}_{\Gamma}- Q_{\Gamma} \tilde{N}_{\Gamma})-M_{\Gamma in}^{-1}(\infty)\in \mathcal{H}_2$, it holds
\[
\langle M_{\Gamma in}^{-1}-M_{\Gamma in}^{-1}(\infty),
M_{\Gamma
out}(\tilde{X}_{\Gamma}- Q_{\Gamma} \tilde{N}_{\Gamma})-M_{\Gamma in}^{-1}(\infty)
\rangle=0.
\]
Hence, (\ref{Inner_factorization1}) is written as follows:
\begin{align}\label{Inner_factorization2}
\hspace{0cm}\left\|\Gamma^{1/2}T_j \gamma_j^{-1}\right\|_2^2
=&\left\|\left[M_{\Gamma out}(\tilde{X}_{\Gamma}- Q_{\Gamma} \tilde{N}_{\Gamma})-M_{\Gamma
in}^{-1}(\infty)\right]e_j\right\|_2^2\nonumber\\ &\hspace{1cm} +\left\|\left[ M_{\Gamma in}^{-1}-M_{\Gamma
in}^{-1}(\infty) \right]e_j \right\|_2^2.
\end{align}
Let
\begin{align}\label{J_Q_Gamma}
&J_{\Gamma}(Q_{\Gamma})=\max_{1\leq j \leq r}\left\{ \left[\left\|\left[M_{\Gamma out}(\tilde{X}_{\Gamma}- Q_{\Gamma} \tilde{N}_{\Gamma})-M_{\Gamma in}^{-1}(\infty)\right]e_j\right\|_2^2 \right.\right. \nonumber\\
&\hspace{0.2cm} \left.\left.+\left\|\left[ M_{\Gamma in}^{-1}-M_{\Gamma
in}^{-1}(\infty) \right]e_j \right\|_2^2\right]\frac{p_j}{1-p_j}, j=1,\cdots, r\right\}.
\end{align}
According to Lemma \ref{lemma_stable}, (\ref{Inner_factorization2}) and (\ref{J_Q_Gamma}), the system is mean-square stabilizible if and only if it holds for some $\Gamma >0$ that
\begin{align}\label{spectral_radius1}
&\min_{Q_{\Gamma}\in \mathcal{RH}_{\infty}}J_{\Gamma}(Q_{\Gamma})<1.
\end{align}
For any given $\Gamma>0$, it is hard to find an analytic optimal solution in minimizing $J_{\Gamma}(Q_{\Gamma})$, in general. To explain the intrinsic constraints caused by interaction between packet dropout probabilities, unstable poles and non-minimum phase zeros in this mean-square stabilization problem, the analytic optimal solution in minimizing $J_{\Gamma}(Q_{\Gamma})$ is studied for the system under Assumption \ref{NMP_zero}.

From Assumption \ref{NMP_zero}, an inner-outer factorization of $\tilde{N}_{\Gamma}$ is given by
$\tilde{N}_{\Gamma}=\tilde{N}_{\Gamma out} {\rm diag}\left\{n_{1,in},\cdots,n_{r,in}\right\}$
where $\tilde{N}_{\Gamma out}$ is an outer of
$\tilde{N}_{\Gamma}$ and ${\displaystyle n_{j,in}=\frac{z-z_j}{z_j^*z-1}}$, $j=1,\cdots, r$ are inner
factors. Thus, from $n^{\sim}_{j,in}n_{j,in}=1$, we obtain that
\begin{align}\label{Inner_factorization3}
&\hspace{0cm}\left\|\left[M_{\Gamma out}(\tilde{X}_{\Gamma}- Q_{\Gamma} \tilde{N}_{\Gamma})-M_{\Gamma
in}^{-1}(\infty)\right]e_j\right\|_2^2\nonumber\\
\hspace{-0.8cm}=&\left\|M_{\Gamma out} Q_{\Gamma} \tilde{N}_{\Gamma out}e_j
-\left[M_{\Gamma out} \tilde{X}_{\Gamma}-M_{\Gamma in}^{-1}(\infty)\right]e_jn_{j,in}^{-1}\right\|_2^2.
\end{align}
Subsequently, it follows from fraction decomposition that
\begin{align}\label{fraction_decomposition}
&\left[M_{\Gamma out} \tilde{X}_{\Gamma}-M_{\Gamma in}^{-1}(\infty)\right]e_jn_{in,j}^{-1}\nonumber\\
&\hspace{-0.4cm}=-\left[M_{\Gamma out}(z_j) \tilde{X}_{\Gamma}(z_j)-M_{\Gamma
in}^{-1}(\infty)\right]e_j\frac{1-z_j^*z_j}{z-z_j}+L_j,
\end{align}
where $L_j$ is the remainder part of this fraction decomposition which belongs to $\mathcal{H}_2$.
Note the fact that
\[
\left[M_{\Gamma out}(z_j) \tilde{X}_{\Gamma}(z_j)-M_{\Gamma
in}^{-1}(\infty)\right]e_j\frac{1-z_j^*z_j}{z-z_j}\in \mathcal{H}_2^{\perp}.
\]
Then, substituting (\ref{fraction_decomposition}) into (\ref{Inner_factorization3}) leads to
\begin{align}\label{Inner_factorization4}
&\left\|\left[M_{\Gamma out}(\tilde{X}_{\Gamma}- Q_{\Gamma} \tilde{N}_{\Gamma})-M_{\Gamma
in}^{-1}(\infty)\right]e_j\right\|_2^2\nonumber\\ =&\left\|M_{\Gamma out} Q_{\Gamma} \tilde{N}_{\Gamma
out}e_j -L_j\right\|_2^2\nonumber\\ &\hspace{0.2cm}+\left\| \left[M_{\Gamma out}(z_j)
\tilde{X}_{\Gamma}(z_j)-M_{\Gamma in}^{-1}(\infty)\right]e_j\frac{1-z_j^*z_j}{z-z_j} \right\|_2^2.
\end{align}

Let $L=\left[L_1\, \cdots \, L_r \right]$. Select
$Q_{\Gamma}= M_{\Gamma out}^{-1} L \tilde{N}_{\Gamma out}^{-1}$.
It is clear from (\ref{Inner_factorization2}) and
(\ref{Inner_factorization4}) that this $Q_{\Gamma}$ minimizes $\left\|\Gamma^{1/2} T_j \gamma_j^{-1}
\right\|_2^2$, $j=1,\cdots, r$ simultaneously. Moreover, it holds that
\begin{align}\label{J_Gamma_min}
\min_{Q_{\Gamma}\in \mathcal{RH}_{\infty}}J_{\Gamma}(Q_{\Gamma})=\max_{1\leq j \leq r}\left\{J_{\Gamma,j}
\frac{p_j}{1-p_j}, \;j=1,\cdots, r\right\}
\end{align}
where
\begin{align}\label{Inner_factorization6}
&\hspace{0cm}J_{\Gamma, j}=\min_{Q_{\Gamma}} \left\|\Gamma^{1/2} T_j \gamma_j^{-1}\right\|_2^2 =\left\|\left[ M_{\Gamma
in}^{-1}-M_{\Gamma in}^{-1}(\infty) \right]e_j \right\|_2^2\nonumber\\
&\hspace{0.6cm}+\left\|
\left[M_{\Gamma out}(z_j) \tilde{X}_{\Gamma}(z_j)-M_{\Gamma
in}^{-1}(\infty)\right]e_j\frac{1-z_j^*z_j}{z-z_j} \right\|_2^2. 
\end{align}
Denote the packet dropout probability vector by $p=(p_1, \cdots, p_r)$ and the mean-square stabilizible region of $p$ to the closed-loop system by ${\mathcal P}$. This mean-square stabilizible region ${\mathcal P}$ is studied in terms of the non-minimum phase zeros and a balanced realization of $M_{\Gamma in}$ given by
\begin{align*}
M_{\Gamma in}=\sys{A_{\Gamma in}}{B_{\Gamma in}}{C_{\Gamma in}}{D_{\Gamma in}}.
\end{align*}
\begin{theorem}\label{lemma_optimal_R_NMP}
Suppose that the plant $G$ satisfies Assumption \ref{NMP_zero}. The system in Fig.~1 is mean-square stabilizable
if and only if the packet dropout probability vector $p=(p_1, \cdots, p_r) \in {\mathcal P}$ and ${\mathcal P}$ is given by
\begin{align}\label{P_mathcal}
{\mathcal P}&=\left\{ p=(p_1,\cdots, p_r)\left|  p_j < \left(e_j^T\Phi_{\Gamma,j}e_j+1\right)^{-1}  \right.\right.\nonumber\\
&\hspace{2cm}  j=1,\cdots, r, \;\;\; \Gamma>0\Big\}
\end{align}
\label{ind_b}
where
\[
\Phi_{\Gamma,j}=D^{*-1}_{\Gamma in}B^{*}_{\Gamma in}N^*_{j, in}(A_{\Gamma in}^{*-1}) N_{j, in}(A_{\Gamma in}^{*-1})B_{\Gamma in}D^{-1}_{\Gamma in}
\]
and
\[
N_{j,in}(A_{\Gamma in}^{*-1})=(z^*_jA^{*-1}_{\Gamma in}-I)(z_jI-A^{*-1}_{\Gamma in})^{-1}.
\]
\end{theorem}
The proof of this theorem is given in Appendix  A.

In general, the mean-square stabilizible region ${\mathcal P}$ given by this theorem is non-convex. Now, a set of convex subregions of ${\mathcal P}$ is studied in terms of diagonal inners $M_{in}$ associated with Wonham decomposition forms of the plant $G\mu$.
\begin{theorem}\label{MIMOstabilization_NMP}
Suppose that the plant $G$ satisfies Assumption \ref{NMP_zero}. Then, the
system in Fig.~1 is mean-square stabilizable
if, for all  $j=1,\cdots, r$, the packet dropout probability $p_j$ in $j$-th channel satisfies:
\begin{align}\label{HR_bound}
p_j \leq \hat{p}_j
\end{align}
where $\hat{p}_j^{-1}=D^{*-1}_{j,in}B^{*}_{j,in}N_{j,in}^*(A^{*-1}_{j,in}) N_{j,in}(A^{*-1}_{j,in})
B_{j,in}D^{-1}_{j,in}+1$.
\end{theorem}
The proof of this theorem is presented in Appendix B. 

It is worth to be mentioned that, in terms of the signal-to-noise ratio, $\log {\displaystyle \frac{1}{p_j}}$ is the counterpart of the channel capacity which is studied in \cite{05E} and \cite{09XXQ}. Instead of considering the minimum channel capacities for the mean-square stabilizability, Theorems \ref{lemma_optimal_R_NMP} and \ref{MIMOstabilization_NMP} describe the mean-square stabilizable region and one of its subregions for the packet dropout probability vector $p=(p_1, \cdots, p_r)$, respectively. In particular, the subregion presented in Theorem \ref{MIMOstabilization_NMP} is a hyper-rectangle determined by a diagonal inner $M_{in}$ which is associated with a given Wonham decomposition form of the plant $G\mu$. Moreover, it has two diagonal vertices which are the origin and $V=(\hat{p}_1, \cdots, \hat{p}_r)$, respectively. The latter is in the boundary of the mean-square stabilizable region $\mathcal{P}$.

Since Wonham decomposition form of the plant $G\mu$ may not be unique, if this is a case, there is a set of such mean-square stabilizable hyper-rectangles for the packet dropout probability vector $p$. Denote the diagonal inner associated with the $s$-th Wonham decomposition form by $M_{s,in}$ and denote its diagonal entries by $m_{s1,in}, \cdots, m_{sr,in}$. Let $\sys{A_{sj,in}}{B_{sj,in}}{C_{sj,in}}{D_{sj,in}}$ be a balance realization of $m_{sj, in}$, $j=1,\cdots, r$. Denote the mean-square stabilizable hyper-rectangle associated with the $s$-th Wonham decomposition form by ${\mathcal P}_s$. Its vertex $V_{s}=(\hat{p}_{s1},\cdots, \hat{p}_{sr})$ is obtained by using Theorem \ref{MIMOstabilization_NMP}. 
\begin{corollary}\label{union_P}
If the packet dropout probability vector $(p_1, \cdots, p_r)$ is in the union of all ${\mathcal P}_s$, i.e.,
\begin{align}\label{union_P_1}
(p_1, \cdots, p_r) \in \underset{s}{\cup} {\mathcal P}_s,
\end{align}
then the networked feedback system in Fig. 1 is mean-square stabilizable.
\end{corollary}
If the plant $G$ has only one Wonham decomposition form, the mean-square stabilizable hyper-rectangles merge to one hyper-rectangle. The equation (\ref{union_P_1}) becomes the necessary and sufficient condition for the mean-square stabilizability of the system. For a SIMO plant $G$, there is only one Wonham decomposition form, the mean-square stabilizable region and  hyper-rectangle studied in Theorems \ref{lemma_optimal_R_NMP} and \ref{MIMOstabilization_NMP}, respectively, degrade to a common interval in one dimension space. In this case, Theorem \ref{MIMOstabilization_NMP} presents a necessary and sufficient condition for the mean-square stabilizability of the system, i.e., $\hat{p}_1$ given by the theorem is the supremum of the packet dropout probability which is allowed for the mean-square stabilizability of the network feedback system. In particular, for a SISO plant with one unstable pole $\lambda_1$ and one non-minimum phase zero $z_1$, this supremum is given by
$$\hat{p}_1=\left[(\lambda_1^2-1)(z_1\lambda_1-1)^2/(z_1-\lambda_1)^2+1\right]^{-1}.$$


Notice the fact that the product $\prod_{j=1}^r p_j$ is the probability with which data packets over all channels are dropped simultaneously. In this work, it is referred to as blocking packet dropout probability. The volume of a hyper-rectangle ${\mathcal P}_s$ is the maximum of the blocking packet dropout probability for all $(p_1,\cdots, p_r) \in {\mathcal P}_s$. Thus, it leads to:
\begin{corollary}
If the blocking packet dropout probability $\prod_{j=1}^r p_j$ of the channels satisfies the inequality
\begin{align}
\prod_{j=1}^r p_j < \max_s  \left\{ \prod_{j}^r \hat{p}_{sj}   \right\},
\end{align}
then, there exists a set of data dropout probabilities $p_1, \cdots, p_r$ with which the networked feedback system in Fig. \ref{setupfeedback} is mean-square stabilizable.
\end{corollary}

Now, we study the case in which the non-minimum phase zeros are void, i.e., the plant is a minimum phase system with relative degree one.
\begin{lemma}\label{lemma_optimal_R}
Suppose that the plant $G$ is minimum phase with relative degree one. Then, for any given $\Gamma >0$, there
exists a $Q$ to jointly minimize $\left\|\Gamma^{1/2} T_j \gamma_j^{-1}\right\|_2^2$, $j=1,\cdots,r$. It
holds that
\begin{align*}
\min_Q \left\|\Gamma^{1/2} T_j \gamma_j^{-1}\right\|_2^2=e_j^TD^{*-1}_{\Gamma in}D^{-1}_{\Gamma in}e_j-1.
\end{align*}
\end{lemma}
The proof of this lemma is given in Appendix C. 
\begin{theorem}\label{MIMOstabilization_MP}
If the plant $G$ is minimum phase with relative degree one and $\lambda_1,\cdots,\lambda_l$ are unstable
poles of the plant, then the closed-loop system is mean-square stabilizible for some packet dropout probabilities $p_1$, $\cdots$, $p_r$ if and only if the blocking packet dropout probability of the channels satisfy:
\begin{align}
\prod_{j=1}^r p_j < (\prod_{i=1}^l|\lambda_i|)^{-1}.
\end{align}
\end{theorem}
The proof of this theorem is given in Appendix D. 

Theorem \ref{MIMOstabilization_MP} presents  the supremum of the blocking packet dropout probability in the mean-square stabilizability of the networked feedback system. Once the blocking packet dropout rate is less than this supremum, the system is mean-square stabilizable by allocating packet dropout probabilities among channels.

\begin{example}
{\rm Suppose that the plant in the networked feedback system shown in Fig 1. is a two-input two-output system. The transfer function of the plant is given as below:}
\begin{align*}
G=\begin{bmatrix} {\displaystyle  \frac{(z-0.25)(z+2)}{z(z-2)(z+1.5)}}  &  {\displaystyle \frac{z-1.5}{z(z+1.5)}}  \\
{\displaystyle \frac{z+2}{z(z-2)} }  &  {\displaystyle \frac{(2z-2.75)(z-1.5)}{z(z-0.25)(z-2.5)}}  \end{bmatrix}.
\end{align*}
\end{example}
Let $p_1$ and $p_2$ be packet dropout probabilities of two channels, respectively. Applying Theorem \ref{lemma_optimal_R_NMP}, we can obtain the mean-square stabilizable region $OV_{11}V_1V_2V_{22}$ in Fig. \ref{H2norm} for the packet dropout probabilities, numerically.

Note the fact that there are two Wonham decomposition forms for the plant. Two diagonal inners associated with these forms are
\[
M_{1,in}={\rm diag} \left\{ \frac{z-2}{2z-1}, \frac{(z+1.5)(z-2.5)}{(-1.5z-1)(2.5z-1)}\right\}
\]
and
\[
M_{2, in}={\rm diag} \left\{ \frac{(z-2)(z+1.5)}{(2z-1)(-1.5z-1)}, \frac{z-2.5}{2.5z-1}\right\}.
\]
The balance realizations of $M_{1, in}$ and $M_{2, in}$ are given  by
\begin{align}\label{Minner1}
M_{1, in}=\sys{\begin{bmatrix}0.5  &  0   &    0    \\
                              0    &  0.4 & -0.683  \\
                              0    &  0   &  -0.667\end{bmatrix}}
{\begin{bmatrix} -0.866  &  0     \\
                  0      &  0.611 \\
                  0      & -0.745\end{bmatrix}}
{\begin{bmatrix} 0.866   &   0     &    0     \\
                 0       &   0.917 &    0.298 \end{bmatrix}}
{\begin{bmatrix}    0.5  &   0     \\
                    0    &  -0.267 \end{bmatrix}  }
\end{align}
and
\begin{align}\label{Minner2}
M_{2, in}=\sys{\begin{bmatrix}0.5 &   0.646 &  0    \\
                              0   &  -0.667 &  0    \\
                              0   &   0     &  0.4  \end{bmatrix}}
{\begin{bmatrix}  0.578 &   0   \\
                  0.745 &   0   \\
                  0     &  -0.917\end{bmatrix}}
{\begin{bmatrix}0.866 & -0.373 &    0     \\
                0     &  0     &    0.817  \end{bmatrix}}
{\begin{bmatrix}    -0.333  &   0    \\
                     0      &   0.4  \end{bmatrix}},\nonumber\\
\end{align}
respectively.
According to Theorem \ref{MIMOstabilization_NMP}, the mean-square stabilizable regions $p_1<\hat{p}_{11}=0.1758$ and $p_2<\hat{p}_{12}=0.0142$ are obtained from the balance realization (\ref{Minner1}) of $M_{1, in}$ for $p_1$ and $p_2$, respectively. Subsequently, the mean-square stabilizable rectangle $OV_{11}V_1V_{12}$ shown in Fig. \ref{H2norm} is obtained for the packet dropout probability vector $(p_1, \ p_2)$. Similarly, the mean-square stabilizable regions $p_1<\hat{p}_{21}=0.0476$ and $p_2<\hat{p}_{22}=0.0246$ are obtained from the balance realization (\ref{Minner2}) of $M_{2, in}$ for $p_1$ and $p_2$, respectively. The mean-square stabilizable rectangle $OV_{21}V_2V_{22}$ shown in Fig. \ref{H2norm} is obtained for the packet dropout probability vector. Two vertices of these rectangles are $V_1=(0.1758, 0.0142)$ and $V_2=(0.0476, 0.0246)$, the areas of the rectangles are $2.50 \times 10^{-3}$, $1.17\times 10^{-3}$, respectively. In other words, the mean-square stabilizable rectangle obtained from $M_{1, in}$ is bounded by the green curve $p_1p_2=2.50\times 10^{-3}$. While, the mean-square stabilizable rectangle obtained from $M_{2, in}$ is bounded by the green curve $p_1p_2=1.17\times 10^{-3}$. The upper bound of the blocking packet dropout probability for mean-square stabilizabilty of the system is $2.50 \times 10^{-3}$. If the plant had only one Wonham decomposition form, these two green curves would merge to one curve and the two rectangles would merge to one rectangle as well.
\begin{figure}
  \centering
  \hspace{-0.5cm}\includegraphics[width=9cm]{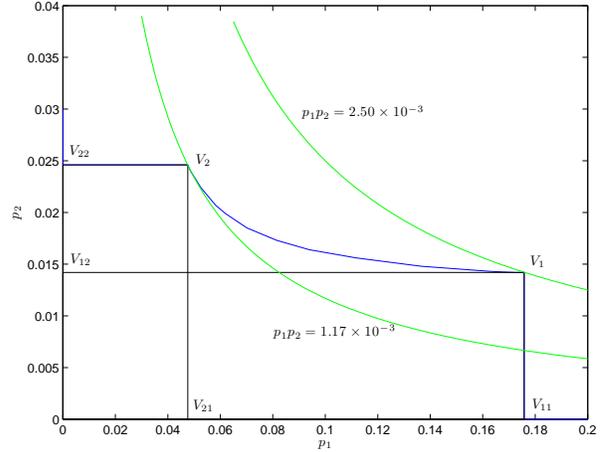}
  \caption{Mean-Square Stabilizable Region for Data Dropout Rate}
  \label{H2norm}
\end{figure}

\section{Conclusion}

This work studies the mean-square stabilizability via output feedback for a networked MIMO feedback system
over several parallel packet dropping communication channels. The admissible region of packet dropout probabilities is discussed in the mean-square stabilizability of a non-minimum phase networked system. The trade-off among these packet dropout probabilities, plant's characteristics and structure in the mean-square stabilizability of the system is presented by an upper bound of blocking packet dropout probability in the region. And then, it is found that, for a minimum phase plant with relative degree one, the supremum of blocking packet dropout probability which is allowed for the mean-square stabilizability is only determined by the product of the plant's unstable poles.

\appendix


\section*{Appendix  A  Proof of Theorem \ref{lemma_optimal_R_NMP}}\label{lemma_optimal_R_NMP_proof}

\renewcommand{\theequation}{A.\arabic{equation}}

We first review three basic results in LTI systems.

\begin{lemma}(see \cite{97FH})\label{lemma10}
For a balanced realization ${\displaystyle
\sys{A_{j,in}}{B_{j,in}}{C_{j,in}}{D_{j,in}}}$ of the inner $m_{j,in}$, it holds that
\begin{align}
\begin{bmatrix}
A_{j, in}^* & C_{j, in}^*\\ B_{j, in}^* & D_{j, in}^*
\end{bmatrix}
\begin{bmatrix}
A_{j, in} & B_{j, in}\\ C_{j, in} & D_{j, in}
\end{bmatrix}
=
\begin{bmatrix}
I & 0\\ 0 & I
\end{bmatrix}
.
\end{align}
\end{lemma}
\begin{lemma}(see e.g. \cite{96ZDG})\label{H2_norm}
Suppose that the system $G(z)=\sys{A}{B}{C}{D}$ is stable and that matrix $P$ is the positive semi-definite
solution to the following Lyapunov equation: 
\[
A^*PA-P+C^*C=0.
\]
Then the ${\mathcal H}_2$-norm $\|G\|_2^2$ of the system is given by
\begin{align*}
\|G\|_2^2
={\rm tr}\left\{B^* P B+D^*D \right\}.
\end{align*}
\end{lemma}
\begin{lemma}\label{inverse}
Suppose that the transfer matrix $G(z)=\sys{A}{B}{C}{D}$ is square and $D$ is invertible. Then the inverse
of G(z) is given by $G^{-1}(z)=\sys{A-BD^{-1}C}{-BD^{-1}}{D^{-1}C}{D^{-1}}.$
\end{lemma}

To prove this theorem, we start from (\ref{J_Gamma_min}) and (\ref{Inner_factorization6}).
It follows from Bezout identity and Assumption \ref{NMP_zero} that
\begin{align}\label{NMP_Bezout}
M_{\Gamma}(z_j)\tilde{X}_{\Gamma}(z_j)e_j=e_j.
\end{align}
Substituting $M_{\Gamma}(z_j)=M_{\Gamma in}(z_j)M_{\Gamma out}(z_j)$ into (\ref{NMP_Bezout}) yields
\begin{align*}
M_{\Gamma out}(z_j)\tilde{X}_{\Gamma}(z_j)e_j=M_{\Gamma in}^{-1}(z_j)e_j.
\end{align*}
Hence, we rewrite (\ref{Inner_factorization6}) as follows:
\begin{align}\label{Inner_factorization7}
&\hspace{-0.5cm}J_{\Gamma, j}=\min_{Q_{\Gamma}} \left\|\Gamma^{1/2} T_j \gamma_j^{-1}\right\|_2^2=\left\|\left[ M_{\Gamma
in}^{-1}-M_{\Gamma in}^{-1}(\infty) \right]e_j \right\|_2^2\nonumber\\ &\hspace{1cm}+\left\| \left[M_{\Gamma
in}^{-1}(z_j) - M_{\Gamma in}^{-1}(\infty)\right]e_j\frac{1-z_j^*z_j}{z-z_j} \right\|_2^2, \nonumber\\
&\hspace{4cm} j=1,\cdots,r.
\end{align}

According to Lemma \ref{lemma_stable} and (\ref{Inner_factorization7}), the system is mean-square stabilizible if and only if it holds for a $\Gamma>0$ that
\begin{align}\label{Inner_factorization8}
J_{\Gamma, j}< \frac{1}{p_j}-1, \quad j=1,\cdots,r.
\end{align}

Now we consider the expression of $J_{\Gamma, j}$, $j=1,\cdots, r$ in terms of the balance realization $\sys{A_{\Gamma in}}{B_{\Gamma in}}{C_{\Gamma in}}{D_{\Gamma in}}$ of $M_{\Gamma in}$.
Since $M_{\Gamma in}$ is an inner, it holds that
\begin{align*}
\left\|\left[ M_{\Gamma in}^{-1}-M_{\Gamma in}^{-1}(\infty) \right]e_j \right\|_2^2
=\left\|\left[ I-M_{\Gamma in}M_{\Gamma in}^{-1}(\infty) \right]e_j \right\|_2^2.
\end{align*}
Now, applying the balanced realization, we have
\begin{align*}
[M_{\Gamma in}M_{\Gamma in}^{-1}(\infty)-I]e_j=C_{\Gamma in}\left(zI-A_{\Gamma in}\right)^{-1}B_{\Gamma in}
D^{-1}_{\Gamma in}e_j.
\end{align*}
According to Lemmas \ref{lemma10} and \ref{H2_norm}, it is obtained that
\begin{align}\label{Cost_MP}
\left\|\left[I-M_{\Gamma in}M_{\Gamma in}^{-1}(\infty) \right]e_j \right\|_2^2
=e_j^T D^{*-1}_{\Gamma in}B^*_{\Gamma in}   B_{\Gamma in} D^{-1}_{\Gamma in}e_j.
\end{align}
On the other hand, it holds that
\begin{align}\label{Cost_NMP}
&\left\| \left[M_{\Gamma in}^{-1}(z_j) - M_{\Gamma in}^{-1}(\infty)\right]e_j\frac{1-z_j^*z_j}{z-z_j}
\right\|_2^2\nonumber\\ =&(z_j^*z_j-1)e^T_j\left[M_{\Gamma in}^{-1}(z_j) - M_{\Gamma
in}^{-1}(\infty)\right]^*\nonumber\\ &\hspace{3cm}\times \left[M_{\Gamma in}^{-1}(z_j) - M_{\Gamma
in}^{-1}(\infty)\right]e_j
\end{align}
where $M_{\Gamma in}^{-1}$ is given by Lemma \ref{inverse} that
\begin{align*}
M_{\Gamma in}^{-1}=\sys{A_{\Gamma in}-B_{\Gamma in}D^{-1}_{\Gamma in}C_{\Gamma in}}{-B_{\Gamma
in}D^{-1}_{\Gamma in}}{D^{-1}_{\Gamma in}C_{\Gamma in}}{D^{-1}_{\Gamma in}}.
\end{align*}
It is verified by Lemma \ref{lemma10} that
$A^{*-1}_{\Gamma in}=A_{\Gamma in}-B_{\Gamma in}D^{-1}_{\Gamma in}C_{\Gamma in}$.
So, we have
\begin{align}\label{Cost_NMP1}
M_{\Gamma in}^{-1}(z_j) - M_{\Gamma in}^{-1}(\infty)=-D^{-1}_{\Gamma in}C_{\Gamma in}(z_jI-A^{*-1}_{\Gamma
in})^{-1}B_{\Gamma in}D^{-1}_{\Gamma in}.
\end{align}
Substituting (\ref{Cost_MP}), (\ref{Cost_NMP}), (\ref{Cost_NMP1}) into (\ref{Inner_factorization7}) leads
to
\begin{align*}
J_{\Gamma, j} =&e_j^TD^{*-1}_{\Gamma
in}B^*_{\Gamma in}(z_j^*I-A^{-1}_{\Gamma in})^{-1}\left[(z_j^*z_j-1)C^*_{\Gamma in}D^{*-1}_{\Gamma
in}D^{-1}_{\Gamma in}C_{\Gamma in}\right.\nonumber\\ &+\left.(z_j^*I-A^{-1}_{\Gamma
in})(z_jI-A^{*-1}_{\Gamma in})
\right](z_jI-A^{*-1}_{\Gamma in})^{-1}B_{\Gamma in}D^{-1}_{\Gamma in}e_j
\end{align*}
It follows from Lemma \ref{lemma10} that
\[
C_{\Gamma in}^*D^{*-1}_{\Gamma in}D^{-1}_{\Gamma in}C_{\Gamma in}+I=A^{-1}_{\Gamma in}A^{*-1}_{\Gamma in}.
\]
Then it turns out that
\begin{align}\label{J_Q_j}
&J_{\Gamma, j} =e_j^TD^{*-1}_{\Gamma in}B^*_{\Gamma
in}(z^*_jI-A^{-1}_{\Gamma in})^{-1}(z_jA^{-1}_{\Gamma in}-I)\nonumber\\
&\hspace{1cm}\times
(z^*_jA^{*-1}_{\Gamma in}-I)(z_jI-A^{*-1}_{\Gamma in})^{-1}B_{\Gamma in}D^{-1}_{\Gamma in}e_j.
\end{align}
Consequently, from (\ref{Inner_factorization8}) and (\ref{J_Q_j}), we obtain that the system is mean-square stabilizible if and only if $p=(p_1,\cdots, p_r) \in {\mathcal P}$.

\section*{Appendix B Proof of Theorem \ref{MIMOstabilization_NMP}}\label{MIMOstabilization_NMP_proof}

\renewcommand{\theequation}{B.\arabic{equation}}
\setcounter{equation}{0}

Since $NM^{-1}$ is an upper triangular coprime factorization (see Section 3), matrices $M$ and $M_{\Gamma}$
are written as follows:
\begin{align*}
M=\begin{bmatrix} m_{11} & m_{12}&\cdots & m_{1r}\\ 0  & m_{22} &  \cdots & m_{2r}\\
   & \cdots  &  \cdots &       \\
0  &  0     &  \cdots & m_{rr}
\end{bmatrix}
,\;\;\; 
M_{\Gamma}=
\begin{bmatrix}
m_{11} &  \hat{m}_{12}   &  \cdots & \hat{m}_{1r} \\ 0 &  m_{22} &  \cdots  & \hat{m}_{2r}  \\
  &   \cdots      &  \cdots  &  \\
0 &   0  &  \cdots &  m_{rr}
\end{bmatrix}
\end{align*}
where $\hat{m}_{ij}=m_{ij}\frac{\gamma_i}{\gamma_j}$, $1\leq i<j
\leq r$.

Let $\Gamma_{\varepsilon}^{1/2}={\rm diag} \left\{1, \varepsilon^{-1}, \cdots, \varepsilon^{-r+1} \right\} $.
\begin{align*}
M_{\Gamma_{\varepsilon}}=
\begin{bmatrix}
m_{11} &  m_{12}\varepsilon  &  \cdots & m_{1r}\varepsilon^{r-2} \\ 0 &  m_{22} &  \cdots  &
m_{2r}\varepsilon^{r-3}  \\
  &   \cdots      &  \cdots  &  \\
0 &   0  &  \cdots &  m_{rr}
\end{bmatrix}.
\end{align*}
As studied in Section \ref{Upper_Tri_CPF}, selecting proper $f_j$, $j=1,\cdots, r$ in Lemma \ref{UT} yields that $m_{11}$, $\cdots$, $m_{rr}$ are inner factors $m_{1, in}$, $\cdots$, $m_{r, in}$. This leads to
$
\underset{\varepsilon \rightarrow 0}\lim M_{\Gamma_{\varepsilon}}={\rm diag}\left\{m_{1,in}, \cdots,
m_{r,in}\right\},
$
i.e., $\underset{\varepsilon \rightarrow 0}\lim M_{\Gamma_{\varepsilon}}$ is an inner.
Let $M_{\Gamma_{\varepsilon}}=M_{\Gamma_{\varepsilon},in}M_{\Gamma_{\varepsilon},out}$ be inner-outer factorization. It holds that
\begin{align}\label{M_epsilon}
\underset{\varepsilon \rightarrow 0}\lim M_{\Gamma_{\varepsilon}, in}={\rm diag}\left\{m_{1,in}, \cdots,
m_{r,in}\right\}.
\end{align}

It is clear that
\begin{align*}
\inf_{\Gamma}\inf_{Q}
\left\|\Gamma^{1/2} T_j\gamma_j^{-1}\right\|_2^2
\leq \lim_{\varepsilon \rightarrow 0}\inf_{Q}
\left\|\Gamma_{\varepsilon}^{1/2}T_j\varepsilon^{j-1}\right\|_2^2.
\end{align*}
From (\ref{Inner_factorization7}) and (\ref{M_epsilon}), it holds that
\begin{align*}
&\hspace{-0.3cm}\lim_{\varepsilon \rightarrow 0}\inf_{Q}
\left\|\Gamma_{\varepsilon}^{1/2} T_j\varepsilon^{j-1}\right\|_2^2 = \left\{
\left\| m_{j, in}^{-1}-m_{j, in}^{-1}(\infty) \right\|_2^2 \right.\nonumber\\
&\hspace{1cm} +\left. \left\| \left[m_{j, in}^{-1}(z_j)-m_{j,
in}^{-1}(\infty)\right]\frac{1-z_j^*z_j}{z-z_j} \right\|_2^2\right\}.
\label{low_b}
\end{align*}
In the light of the proof for Theorem \ref{lemma_optimal_R_NMP}, we can see that
\begin{align*}
&\hspace{0cm}
\left\| m_{j, in}^{-1}-m_{j, in}^{-1}(\infty) \right\|_2^2 + \left\| \left[m_{j, in}^{-1}(z_j)-m_{j,
in}^{-1}(\infty)\right]\frac{1-z_j^*z_j}{z-z_j} \right\|_2^2\nonumber\\
=&D^{*-1}_{j,in}B^{*}_{j,in}N_{j,in}^{*}(A^{*-1}_{j,in}) N_{j,in}(A^{*-1}_{j,in}) B_{j,in}D^{-1}_{j,in}.
\end{align*}

If it holds for $ j=1,\cdots,r$ that
\begin{align*}
&\frac{p_j}{1-p_j}D^{*-1}_{j,in}B^{*}_{j,in}N_{j,in}^*(A_{j,in}) N_{j,in}(A_{j,in})
B_{j,in}D^{-1}_{j,in}<1,
\end{align*}
i.e., the inequality (\ref{HR_bound}) holds for $r=1,\cdots, r$.
Then,
we can design $Q$ such that the inequalities hold
\begin{align*}
&\hspace{0.5cm}\inf_{\Gamma}\inf_{Q}\frac{p_j}{1-p_j}
\left\|\Gamma^{1/2} T_j\gamma_j^{-1}\right\|_2^2<1,\quad
j=1,\cdots,r.
\end{align*}
Therefore, the system is mean-square stabilizible.

\section*{Appendix C Proof of Lemma \ref{lemma_optimal_R}}\label{lemma_optimal_R_proof}

\renewcommand{\theequation}{C.\arabic{equation}}
\setcounter{equation}{0}

It is shown in the proof of Theorem \ref{lemma_optimal_R_NMP} that
\begin{align*}
\left\|\Gamma^{1/2} T_j \gamma_j^{-1}\right\|_2^2
=&\left\|\left[M_{\Gamma out}(\tilde{X}_{\Gamma}- Q_{\Gamma} \tilde{N}_{\Gamma})-M_{\Gamma
in}^{-1}(\infty)\right]e_j\right\|_2^2\nonumber\\ &+\left\|\left[ M_{\Gamma in}^{-1}-M_{\Gamma
in}^{-1}(\infty) \right]e_j \right\|_2^2.
\end{align*}
Since the plant $G$ is minimum phase with relative degree one, we can design $Q_{\Gamma}$ such that
\[
M_{\Gamma out}(\tilde{X}_{\Gamma}- Q_{\Gamma} \tilde{N}_{\Gamma})-M_{\Gamma in}^{-1}(\infty)=0.
\]
Thus, it holds for $j=1,\cdots,r$ that
\begin{align}\label{Min_Tj}
\min_Q\left\|\Gamma^{1/2} T_j \gamma_j^{-1}\right\|_2^2
=\left\|\left[ M_{\Gamma in}^{-1}-M_{\Gamma in}^{-1}(\infty) \right]e_j \right\|_2^2.
\end{align}
Applying (\ref{Cost_MP}), we write (\ref{Min_Tj}) as follows:
\begin{align}\label{Min_Tj1}
\hspace{-0.2cm}\min_Q\left\|\Gamma^{1/2} T_j \gamma_j^{-1}\right\|_2^2
=e_j^T D_{\Gamma in}^{*-1}B_{\Gamma in}^*B_{\Gamma in}D_{\Gamma in}^{-1}e_j, j=1,\cdots,r.
\end{align}
Taking into account of the fact given by Lemma \ref{lemma10}
\[
B_{\Gamma in}^*B_{\Gamma in}+D_{\Gamma in}^*D_{\Gamma in}=I,
\]
we rewrite (\ref{Min_Tj1}) as follows:
\begin{align*}
\min_Q\left\|\Gamma^{1/2} T_j \gamma_j^{-1}\right\|_2^2
=e_j^T D_{\Gamma in}^{*-1}D_{\Gamma in}^{-1}e_j-1, j=1,\cdots,r.
\end{align*}

\section*{Appendix D Proof of Theorem \ref{MIMOstabilization_MP}}\label{MIMOstabilization_MP_proof}

\renewcommand{\theequation}{D.\arabic{equation}}
\setcounter{equation}{0}

According to Lemmas \ref{lemma_stable} and \ref{lemma_optimal_R}, the system is mean-square stabilizible if
and only if it holds for some $\Gamma>0$ that
\begin{align}\label{SNR}
e_j^T D_{\Gamma in}^{*-1}D_{\Gamma in}^{-1}e_j-1<\frac{1-p_j}{p_j}, j=1,\cdots,r.
\end{align}
Thus, the $j$-th channel's packet dropout probability $p_j$ satisfies the inequality for
the $\Gamma>0$ as below:
\begin{align}\label{j_Capacity}
e_j^TD_{\Gamma in}^{*-1}D_{\Gamma in}^{-1}e_j<\frac{1}{p_j}.
\end{align}
Subsequently, it holds for the blocking packet dropout probability of the channel that
\begin{align}\label{low_C}
\prod_{j=1}^r \frac{1}{p_j} > \prod_{j=1}^r
e_j^TD_{\Gamma in}^{*-1}D_{\Gamma in}^{-1}e_j
\geq  {\rm det}\left(D_{\Gamma in}^{*-1}D_{\Gamma
in}^{-1}\right)
\end{align}
where the second inequality follows from a property of a positive definite matrix (see \cite{85HJ}).

Since the plant $G$ has $l$ unstable poles, the inner $M_{\Gamma in}$ has $l$ factors $M_{\Gamma in, 1}$,
$\cdots$, $M_{\Gamma in, l}$, i.e., $M_{\Gamma in}=M_{\Gamma in, 1} \times \cdots \times M_{\Gamma in, l}$.
Each of these factors is associated with an unstable pole $\lambda_i$ and is given by
\begin{align*}
M_{\Gamma in, i}=U_iU_i^*+\eta_i\eta_i^*\frac{z-\lambda_i}{\lambda_i^*z-1}
\end{align*}
where $\eta_i$ is direction vector of $\lambda_i$ and $\begin{bmatrix}U_i& \eta_i\end{bmatrix}
\begin{bmatrix}U_i^*\\ \eta_i^*\end{bmatrix}=I$.
Hence, it holds for the balanced realization of $M_{\Gamma_{in}}$ that
\begin{align}\label{lambda_i}
D_{\Gamma in}=\prod_{i=1}^n\begin{bmatrix}U_i& \eta_i\end{bmatrix}
\begin{bmatrix}I& 0\\ 0  & {\displaystyle \frac{1}{\lambda_i^*}}\end{bmatrix}
\begin{bmatrix}U_i^*\\ \eta_i^*\end{bmatrix}.
\end{align}
Substituting (\ref{lambda_i}) into (\ref{low_C}) leads to
\begin{align}\label{total_Capacity}
\prod_{j=1}^r p_j < \prod_{i=1}^l|\lambda_i|^{-1}.
\end{align}
On the other hand, if the blocking packet dropout probability satisfies (\ref{total_Capacity}), it is always possible to find a
set of  $p_1,\cdots, p_r$ so that the inequalities hold:
\begin{align}\label{j_Capcity_decouple}
p_j < \prod_{i=1}^{l_j}|\lambda_{j i}|^{-1}, \quad
j=1,\cdots,r
\end{align}
where $\lambda_{j i}$, $i=1,\cdots, l_j$ are unstable poles of
$A_j$.

Let $\Gamma^{1/2}=\left\{1, \varepsilon, \cdots, \varepsilon^{r-1}\right\}$. In the light of the proof for Theorem
\ref{MIMOstabilization_NMP}, we can design a mean-square stabilizing controller for the system when some
$p_1,\cdots, p_r$ satisfy the inequalities (\ref{j_Capcity_decouple}), respectively,
and $\varepsilon \rightarrow 0$.



\end{document}